\documentclass[%
 reprint,
 amsmath,amssymb,
 aps,
]{revtex4-2}

\usepackage{graphicx}
\usepackage{dcolumn}
\usepackage{bm}

\begin{document}

\preprint{APS/123-QED}

\title{Orientation-Dependent Ion Acceleration from Laser-Irradiated Rectangular Nanorings}

\author{Xiaohui Gao}
  \email{gaoxh@utexas.edu}
\affiliation{Department of Physics, Shaoxing University, Shaoxing, Zhejiang 312000, China}

\date{\today}

\begin{abstract}
Laser-driven ion acceleration from nanostructured targets offers a promising route to compact, high-energy ion sources. In this work, we demonstrate through particle-in-cell simulations that rectangular nanoring targets significantly enhance energy absorption and increase the cutoff energy of laser-accelerated ions. The nanoring geometry enables strong field confinement within its hollow core when optimally oriented relative to the laser polarization, leading to hotter electron populations and more robust sheath acceleration. These results demonstrate that rectangular nanorings offer a versatile platform for controlling laser-plasma interactions at solid densities and advancing compact, high-repetition-rate particle sources. 
\end{abstract}

\maketitle
\section{Introduction}
Laser-accelerated ions have emerged as a versatile particle source with diverse applications, ranging from microscale fusion~\cite{Curtis2018NC} to medical therapies~\cite{Linz2016PRAAB} and materials science~\cite{Barberio2018NC}. A central objective in advancing these applications is increasing the energy of the produced ion beams. This goal has driven sustained progress in the field, with proton energies recently exceeding 150~MeV~\cite{Ziegler2024NP}. One direct pathway to enhanced ion energies is increasing the driving laser intensity. However, given the practical ceiling on laser intensity near $10^{22}$~W/cm$^2$, as well as constraints on cost and repetition rate, tailoring the target morphology at the micro- and nanoscale offers an attractive alternative~\cite{Shou2025NC,Tochitsky2025SR}. By exploiting geometry and composition, such targets enable efficient ion acceleration with compact, Hz-level, affordable laser systems at moderately relativistic intensities, opening pathways to practical applications such as compact neutron sources~\cite{Knight2024,GozhevPoP2024,Kemp2019NC}.

Consequently, a wide variety of target morphologies, beyond conventional nanofilms~\cite{Macchi2013}, have been investigated, including clusters~\cite{Ditmire1999N,Jinno2022SR,Won2023NF}, droplets~\cite{DiLucchio2015PP,Becker2019SR}, micro-bars~\cite{Elkind2025CP}, microtapes~\cite{ShenPRX2021}, nanotube arrays~\cite{ToscaPRR2024}, and nanowire structures~\cite{Rocca2024O}. The most extensively studied acceleration mechanism, target normal sheath acceleration (TNSA), relies on charge separation driven by laser-generated hot electrons. A related mechanism underpins the expansion of spherical clusters, where an ambipolar field accelerates ions outward~\cite{DiLucchio2015PP}. Within this landscape, mass-limited targets offer inherent advantages by minimizing energy losses and enhancing coupling efficiency~\cite{Limpouch2008}. Yet, research on such targets has predominantly focused on spherical geometries~\cite{Hilz2018NC} or thin films~\cite{Shou2025NC}; other geometries may offer superior performance by enabling greater control over the laser-plasma interaction.

Nanophotonic studies have demonstrated that subwavelength rectangular apertures~\cite{GarciaVidal2010RoMP, GarciaVidal2005PRL} can support efficient field enhancement within their hollow volumes. This enhancement depends sensitively on the aperture's aspect ratio and the orientation of the incident laser polarization. Since hot electrons are generated via the Brunel mechanism, their temperature scales with the local field strength. An enhanced internal field therefore produces a hotter electron population, which in turn drives more efficient ion acceleration. Beyond ion acceleration, such hot electrons are also central to other high-field processes, including x-ray generation via bremsstrahlung, betatron radiation, Thomson scattering, or K$_\alpha$ emission~\cite{Bagchi2011PP,Pan2024PoP,Shou2023NP,Nishikawa2004APB,Chakravarty2012JAP}.

In this paper, we employ particle-in-cell (PIC) simulations to investigate ion acceleration from laser-irradiated rectangular nanorings. We demonstrate that the nanoring's orientation relative to the laser polarization profoundly influences field confinement and electron dynamics, leading to substantial increases in maximum ion energy. With continued advances in nanofabrication, freestanding nanofilms incorporating nanoring structures, such as nanomeshes, are becoming feasible. Such tailored targets offer a promising route toward optimized, compact ion sources and could enable next-generation, laser-driven neutron sources via nuclear reactions in deuterated materials.

\section{Ion acceleration in rectangular nanorings}
\begin{figure}[htbp]
\centering
\includegraphics[width=0.35\textwidth]{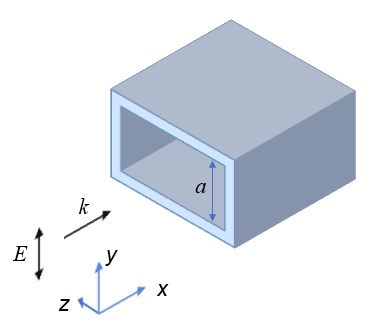}
\caption{Schematic of the nanoring geometry. The nanoring has an inner hollow rectangular region and an outer rectangular profile.}
\label{fig_sche} 
\end{figure}

The interaction geometry is illustrated in Fig.~\ref{fig_sche}. A laser pulse propagates along the $x$ direction with linear polarization along the $y$ axis. The nanoring target has a length $l$ along the laser propagation direction. It features an inner hollow region of rectangular cross section, with dimensions $a$ along the $y$ axis and $b$ along the $z$ axis. The target has an outer rectangular geometry with a uniform shell thickness $t$. In the following, we denote different nanoring geometries by their inner dimensions $a \times b$.

For ion acceleration studies, we consider a CH target composed of low-density polyethylene [(C$_2$H$_4$)$_n$]. The mass density is $0.923$~g/cm$^3$, corresponding to a hydrogen atom number density of $46n_c$ and a carbon atom number density of $23n_c$, where $n_c$ is the critical density for the laser wavelength $\lambda_0 = 800$~nm.

The simulations are performed using the 3D PIC code \textsc{Smilei}~\cite{Derouillat2018CPC} with a spatial resolution of $\lambda_0/80$ and a temporal resolution of $T_0/145$, where $\lambda_0$ is the laser wavelength and $T_0$ is the period. The simulation box spans $8\lambda_0 \times 3\lambda_0 \times 3\lambda_0$, and the nanoring sits at the center of the box, i.e., at $(x,y,z)=(0,0,0)$. Each cell contains 64 neutral macroparticles of hydrogen and carbon. Electrons are generated via field ionization, and collisional ionization is not enabled. Periodic boundary conditions for both fields and particles are applied in the transverse directions. In the $x$ direction, Silver–M\"uller absorbing boundary conditions are used for the fields, while particles exiting the simulation domain are removed.

In all simulations, a linearly polarized laser pulse with an intensity of $2\times10^{18}$~W/cm$^2$ (normalized amplitude $a_0 \approx 0.96$) is injected from the $x$-minimum boundary, and the peak of the pulse reaches $x=0$ at $t=0$ in the absence of the target. The pulse has a Gaussian temporal profile with a full-width at half-maximum of two optical cycles and zero-carrier envelope phase. The shell thickness is fixed at $t = 40$~nm, and the nanoring length is $l = 0.4$~$\mu$m. These parameters, along with the short pulse duration, are chosen primarily to reduce computational cost and do not represent physical limitations of the system.

\begin{figure}[htbp]
\centering
\includegraphics[width=0.40\textwidth]{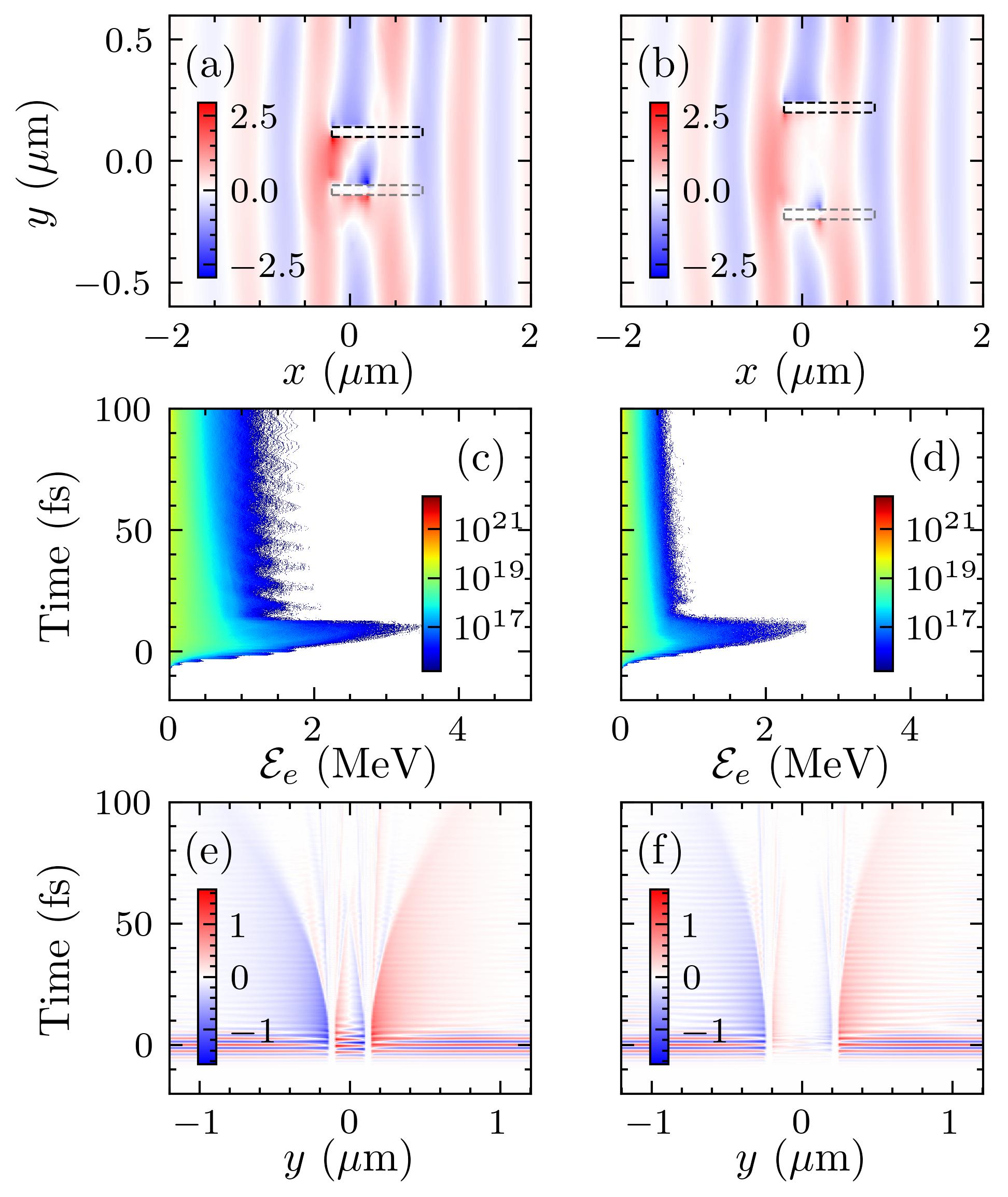}
\caption{Comparison of simulation results for two target geometries: $200$~nm$\times400$~nm ($S_\parallel$, left column) and $400$~nm$\times200$~nm ($S_\perp$, right column).  
(a,b) Instantaneous electric field $E_y$ in the $x$--$y$ plane at $z=0$ and $t = 1.66$\,fs. The color scale is symmetric and normalized to the same maximum absolute value for both panels; the dashed rectangles outline the target positions. (c,d) Streak images of the electron energy spectra: $dN_e/d\mathcal{E}_e$ (cm$^{-3}$\,eV$^{-1}$) versus electron energy $\mathcal{E}_e$ (MeV) and time (fs). The color scale is logarithmic and identical for both cases. (e,f) Spatio-temporal evolution of $E_y$ at the polarization axis [$(x,z) =(0,0)$]: transverse profiles along $y$ ($\mu$m) as a function of time. The field is normalized to the incident amplitude $E_0$ and is identical for both cases.
}
\label{fig_cmp} 
\end{figure}

Figure~\ref{fig_cmp} shows the interaction for two orientations of the same nanoring: $200$~nm$\times400$~nm ($S_\parallel$), where the short side ($200$~nm) is parallel to the laser polarization, and $400$~nm$\times200$~nm ($S_\perp$), where the short side is perpendicular to the polarization. The instantaneous field distributions in the $x$--$y$ plane for $z=0$ and $t=1.66$~fs in Figs.~\ref{fig_cmp}(a) and \ref{fig_cmp}(b) reveal a clear contrast: $S_\parallel$ exhibits significant field enhancement within the hollow aperture, whereas $S_\perp$ shows field suppression. In both cases, field enhancement is also observed at the outer boundaries due to the lightning-rod effect at the sharp corners and polarization charges accumulated on surfaces. Since the electron quiver energy scales as $E^2$, where $E$ is the laser field amplitude, the higher field in $S_\parallel$ produces a hotter electron population, as evident in the electron energy spectra shown in Figs.~\ref{fig_cmp}(c) and \ref{fig_cmp}(d). 
Figures~\ref{fig_cmp}(e) and \ref{fig_cmp}(f) display the spatio-temporal evolution of $E_y$ along the central axis, where a clear sheath field develops as electrons exit the target. The sheath field for $S_\parallel$ is both stronger and propagates faster than for $S_\perp$, consistent with the higher electron temperatures observed.

\begin{figure}[htbp]
\centering
\includegraphics[width=0.40\textwidth]{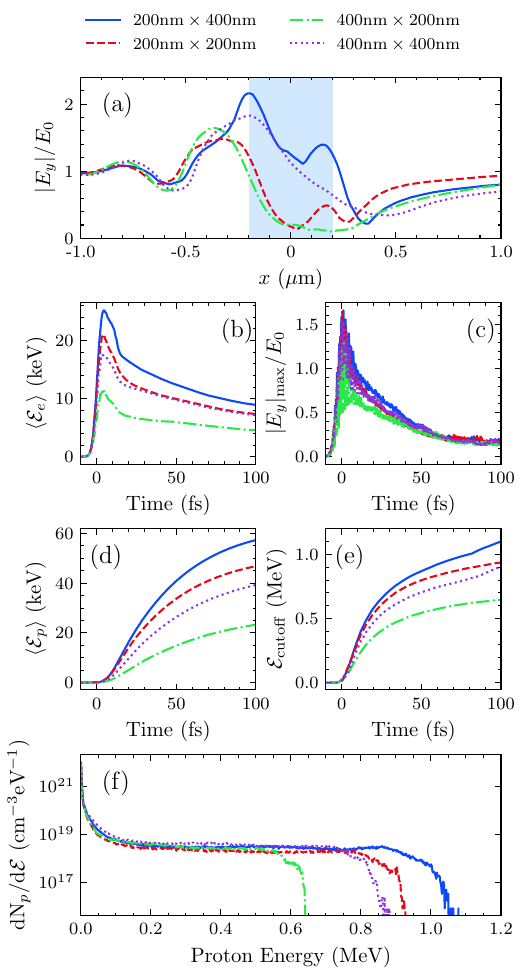}
\caption{Proton acceleration from rectangular plastic nanorings of varying geometry and orientation. Four configurations are compared: $200$~nm$\times400$~nm ($S_\parallel$, blue solid), $400$~nm$\times200$~nm ($S_\perp$, green dash-dotted), $200\times200$~nm square (red dashed), and $400\times400$~nm square (purple dotted). (a) Maximum normalized electric field amplitude $|E_y|/E_0$ along the laser axis ($y=0$) as a function of $x$. The shaded region ($-0.2\,\mu\mathrm{m}<x<0.2\,\mu\mathrm{m}$) indicates the position inside the nanoring. (b) Average electron kinetic energy $\langle \mathcal{E}_e \rangle$ versus time. (c) Peak electric field amplitude $|E_y|$ versus time. (d) Average proton kinetic energy $\langle \mathcal{E}_p \rangle$ versus time. (e) Proton cutoff energy $\mathcal{E}_p^{\text{max}}$ versus time. (f) Proton energy spectra $dN_p/d\mathcal{E}_p$ versus proton energy at $t = 95.2$~fs.}
\label{fig_rect} 
\end{figure}

Figure~\ref{fig_rect} presents the ion acceleration results for the four target geometries. Figure~\ref{fig_rect}(a) shows the field enhancement factor along the laser axis, while Figs.~\ref{fig_rect}(b)--\ref{fig_rect}(e) display the time evolution of the average electron energy, peak sheath field, average proton energy, and proton cutoff energy, respectively. Figure~\ref{fig_rect}(f) shows the proton energy spectra at $t = 95.2$~fs. The $S_\parallel$ configuration ($200$~nm$\times400$~nm, blue solid curve) achieves superior performance across all metrics, with a proton cutoff energy of approximately $1.1$~MeV. In contrast, the $S_\perp$ configuration ($400\times200$~nm, green dash-dotted) reaches only $0.7$~MeV, while the square nanorings (red dashed and purple dotted) exhibit intermediate behavior. Notably, although the interior field enhancement differs considerably between the two square geometries, the time evolution of the average electron energy is remarkably similar. This suggests that the spatially averaged field over the effective interaction volume is comparable, as field enhancement at the exterior boundaries also contributes. Throughout the temporal evolution shown in Figs.~\ref{fig_rect}(b)--\ref{fig_rect}(f), a clear correlation emerges: higher electron temperature leads to a stronger sheath field, which in turn drives higher average and cutoff proton energies.

\begin{figure}[htbp]
\centering
\includegraphics[width=0.40\textwidth]{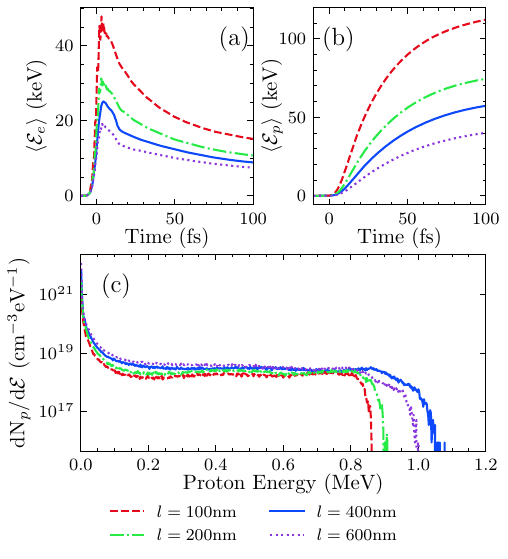}
\caption{Dependence of proton acceleration on nanoring length $l$. Four geometries with different lengths are compared, all with inner cross section $200$~nm$\times400$~nm and shell thickness $40$~nm. (a) Average electron kinetic energy $\langle \mathcal{E}_e \rangle$ versus time. (b) Average proton kinetic energy $\langle \mathcal{E}_p \rangle$ versus time. (c) Proton energy spectra $dN_p/d\mathcal{E}_p$ at $t = 95.2$~fs.}
\label{fig_L} 
\end{figure}

Figure~\ref{fig_L} examines the influence of the nanoring length $l$ along the laser propagation direction. The average energies of electrons and protons decrease with increasing length, as shown in Figs.~\ref{fig_L}(a) and \ref{fig_L}(b). However, $l = 400$~nm yields the highest cutoff proton energy at approximately $1.1$~MeV [Fig.~\ref{fig_L}(c)]. Interestingly, while the $l = 100$~nm case exhibits the highest average electron and proton kinetic energies [Figs.~\ref{fig_L}(a) and (b)], its proton cutoff energy is lower. This apparent contradiction may be due to phase matching between the expanding ion front and the propagating sheath field: with an optimal length, the nanoring allows ions to remain in phase with the accelerating field over an extended distance, converting thermal energy into directed kinetic energy more effectively.

\begin{figure}[htbp]
\centering
\includegraphics[width=0.40\textwidth]{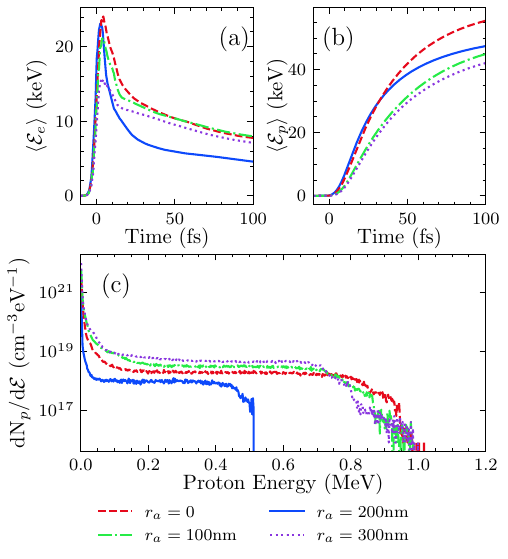}
\caption{Dependence of proton acceleration on inner radius $r_a$ for circular nanorings. Four geometries are compared, with fixed length $l = 400$~nm and shell thickness $40$~nm. $r_a = 0$ corresponds to a solid rod (no hollow core). (a) Average electron kinetic energy $\langle \mathcal{E}_e \rangle$ versus time. (b) Average proton kinetic energy $\langle \mathcal{E}_p \rangle$ versus time. (c) Proton energy spectra $dN_p/d\mathcal{E}_p$ at $t = 95.2$~fs.}
\label{fig_R} 
\end{figure}

Figure~\ref{fig_R} presents results for circular nanorings with varying inner radius $r_a$, all having the same length ($l = 400$~nm) and shell thickness ($40$~nm) as the rectangular cases. Figures~\ref{fig_R}(a) and \ref{fig_R}(b) show the time evolution of the average electron kinetic energy and the average proton kinetic energy, respectively. Figure~\ref{fig_R}(c) displays the proton energy spectra at $t = 95.2$~fs. The $r_a = 0$ case corresponds to a solid rod with no hollow core, serving as a reference. For all hollow nanorings ($r_a > 0$), the average electron energy and proton cutoff energy are remarkably similar despite differences in the internal field structure~\cite{Gao2026PP}, probably due to similarity of spatially averaged field over the effective interaction volume. However, two key distinctions emerge when comparing these results to the rectangular $S_\parallel$ case. First, the cutoff energy for circular nanorings ($\approx 1.0$~MeV) remains below the $1.1$~MeV achieved with the optimally oriented rectangular geometry. Second, the spectral cutoff is more gradual, indicating a lower number of high-energy ions compared to the steeper drop observed for $S_\parallel$.

The solid rod performs significantly worse, with a cutoff energy of approximately $0.5$~MeV. This stark contrast highlights a key advantage of hollow targets: the presence of inner surfaces exposed to vacuum enables efficient vacuum heating and electron extraction from both inner and outer walls. While nanotube arrays have been extensively studied for ion acceleration, their performance typically relies on achieving near-critical-density plasma conditions~\cite{ToscaPRR2024}. Our results demonstrate that hollow nanostructures can enhance acceleration at solid densities through purely geometric control of the local field. Among the geometries considered, the rectangular nanoring with optimal orientation ($S_\parallel$) yields the highest cutoff energy and the steepest spectral slope, confirming that shape optimization beyond simple circular symmetry offers additional benefits for ion acceleration.

\section{Neutron generation}
\begin{figure}[htbp]
\centering
\includegraphics[width=0.40\textwidth]{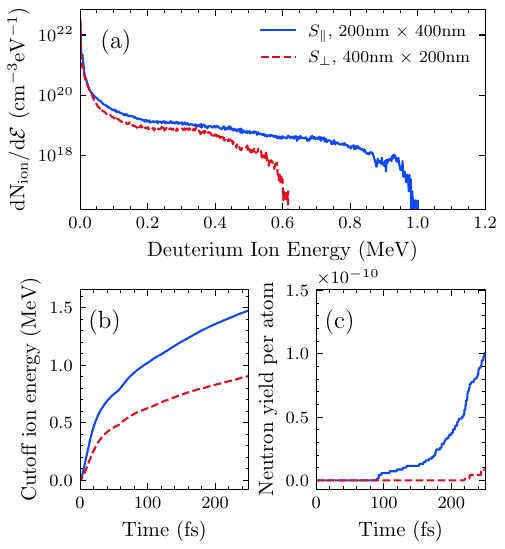}
\caption{Comparison of fusion-relevant quantities for two target orientations: $200$~nm$\times400$~nm ($S_\parallel$, blue solid) and $400$~nm$\times200$~nm ($S_\perp$, red dashed). (a) Ion energy spectra at $t=95.2$~fs, showing the number of ions per unit energy (cm$^{-3}$\,eV$^{-1}$) as a function of energy (MeV). (b) Cutoff ion energy (MeV) as a function of time (fs). (c) Fusion yield per atom versus time (fs). }
\label{fig_N} 
\end{figure}
 To assess the potential of these nanostructured targets for neutron generation, we replace the CH$_2$ composition with CD$_2$ (deuterated polyethylene) while maintaining the same atomic density. The deuterium mass is set to $3870.5\,m_e$. A binary collision module is activated between deuterium ions, with reaction products defined as helium-3 and neutrons according to the D$(d,n)^3$He fusion channel. To enhance the statistical representation of these rare events within the PIC framework, a reaction multiplier of $10^6$ is applied; the physical yield is subsequently recovered by dividing by this factor. The simulation box dimensions are reduced to $8\lambda_0 \times 1.5\lambda_0 \times 1.5\lambda_0$, and periodic boundary conditions in the transverse directions allow ions that exit one side to re-enter from the opposite side. This effectively increases the probability of high-relative-velocity collisions, thereby enhancing the observable reaction rate. The results are shown in Fig.~\ref{fig_N}. Consistent with the proton acceleration trends, the $S_\parallel$ orientation yields a higher cutoff ion energy [Fig.~\ref{fig_N}(b)] and a correspondingly higher fusion yield [Fig.~\ref{fig_N}(c)] compared to $S_\perp$. This demonstrates that geometric optimization for ion acceleration directly translates to enhanced performance in fusion-relevant applications.

The simulations presented here employ a moderately relativistic intensity of $2\times10^{18}$~W/cm$^2$ and a two-cycle pulse. Increasing the intensity and pulse duration would further enhance ion energies. Using a lower intensity regime offers distinct practical advantages: it enables operation at higher repetition rates with more affordable laser systems and relaxes the stringent contrast requirements needed to prevent target pre-ionization by the laser prepulse. Such mass-limited targets with rectangular nanoring features are experimentally feasible; for instance, they could be realized as a $\mu$m-sized nanomesh supported by a thin stalk. Thus the rectangular nanoring geometry is a practically viable candidate for next-generation, laser-driven neutron sources.

\section{Conclusion}
In summary, we have shown that rectangular nanoring targets can substantially enhance proton acceleration compared to solid or circular hollow targets. The orientation of the nanoring relative to the laser polarization critically determines the degree of internal field enhancement, which in turn governs electron heating and the resulting sheath field. Optimal performance is achieved when the short axis of the rectangular aperture aligns with the laser polarization, yielding higher electron temperatures and proton cutoff energies. This geometric control over the local field enables efficient ion acceleration even at solid densities, without requiring near-critical plasma conditions. When applied to deuterated targets, the same configuration enhances fusion neutron yield, demonstrating direct relevance to compact neutron source development. These results underscore the potential of tailored nanostructures for advancing laser-driven particle sources.

\begin{acknowledgments}
This work was supported by the Natural Science Foundation of Zhejiang Province (LY19A040005).
\end{acknowledgments}

\section*{Data Availability Statement}
The data that support the findings of this study are available from the corresponding author upon reasonable request.

%
\end{document}